\documentclass[useAMS,usenatbib]{mn2e}
\usepackage{amsmath}
\usepackage{aas_macros}
\usepackage[latin1]{inputenc}
\usepackage{graphicx}
\usepackage{txfonts}
\usepackage{subfigure}
\newcommand{\ttt}{\times 10^}
\newcommand{\source}{\mbox{1 ES 1218+30.4 }}

\title[Spectral modelling of \source]{Spectral modelling of \source}
\author[M. R\"uger, F. Spanier and K. Mannheim]{M. R\"uger$^{1}$\thanks{E-mail:
mlrueger@astro.uni-wuerzburg.de}, F. Spanier$^{1}$ and K. Mannheim $^{1}$
\\
$^{1}$Lehrstuhl f\"ur Astronomie, Universit\"at W\"urzburg,
	Am Hubland, D-97074 W\"urzburg\\}
\begin{document}

\date{to appear in MNRAS}
\pagerange{\pageref{firstpage}--\pageref{lastpage}} \pubyear{2002}

\maketitle

\label{firstpage}

\begin{abstract}
We employ a time-dependent synchrotron-self-Compton code for modeling contemporaneous
multiwavelength data of the blazar \source The input parameters of the model
are used to infer physical parameters of the emitting region. An acceptable fit to the data is
obtained by taking into account a stellar emission component in the optical regime due to the host
galaxy. The physical parameters inferred from the fit are in line with particle acceleration due
to the Fermi mechanism providing s = 2.1 spectra.
From the properties of the host galaxy in the optical, we estimate the central black hole mass
and thus confirm that the jet power injected into the emission region is in the sub-Eddington regime, as
expected for BL Lacertae objects.

\end{abstract}


\section{Introduction}

Among the class of active galactic nuclei (AGN), blazars are special in
showing a spectral energy distribution (SED) that is strongly dominated by nonthermal
emission across a wide range of wavelengths, from radio waves to gamma rays,
and rapid, large-amplitude variability.   Presumably, these characteristics
are due to a relativistic jet emitted at a small angle to the line-of-sight,
emitting Doppler-boosted synchrotron and inverse-Compton radiation and thus
washing out to a variable extent the emission from the accretion flow and host galaxy.

The synchrotron and inverse-Compton emission could either result from primary
accelerated electrons, from accelerated protons, or from secondary electrons arising in
electromagnetic cascades initiated by pion and pair production \citep[e.g.][]{mannheim93}.

The high peaked BL Lacs (HBLs) show a peak in their SED in the X-ray regime, suggesting
that an inverse-Compton peak should occur at correspondingly high gamma-ray energies.
In fact, a large fraction of the known nearby HBLs have already been
discovered with Cherenkov telescopes, such as H.E.S.S., MAGIC, and VERITAS.

In those cases where the blazars have been detected at gamma-ray energies,
the SED shows two bumps, one at infrared-to-X-ray energies, and the other
at gamma ray energies.  So far, the reasons for the variation of the peak energies
are unknown, and their relation with fundamental parameters of the central
engine, the black hole mass, spin, and accretion rate, are far from settled.

Models of the radiation processes behind the blazar emission are a key issue
to improve our understanding of blazars. The diagnosis of the observed spectra using radiation models allows to
infer the physical conditions prevailing in them and to discern their
relation with fundamental parameters.  However, without a prescription of
the relativistic particle spectra, the models effectively map radiation
onto particle spectra, and the error in the inferred physical conditions
is correspondingly large.  On the other hand, using theoretically favoured
particle spectra most strongly constrains the physical conditions prevailing
in the sources, and this may eventually lead to a consistent physical explanation.

The HBL \source has been
discovered as a candidate BL Lac object on the basis of its X-ray emission and
has been identified with the X-ray source \mbox{2A 1219+30.5} \citep{wilson79,
ledden81}. For the first time, \source  has been observed at  VHE energies using the MAGIC telescope
 in January 2005 \citep{albert06}. Simultaneous optical data has been provided by
 the KVA telescope on La Palma. Latest TeV data have been provided by the VERITAS telescope \citep{veritas09}.

Here, we present the kinetic equation and
numerical code describing the synchrotron-self-Compton
emission (Sect. \ref{sec:model}). The particular emphasis lies on an accurate treatment
of the Klein-Nishina-turnover which is important at very high
gamma ray energies. Previous works (c.f. \citet{boettcher2002} and references in there) provided good results in the scope of HBL modelling. Our approach is quite similar to this models except for details in the treatment of the particular processes. External Compton effects are left out in order to use a minimum number 
of parameters, which are sufficient for dealing with HBLs. In Sect.~\ref{sec:results} we apply our code to \source  referring
in particular to the MAGIC and VERITAS data and give a set of physical parameters for the most
acceptable fit.  Finally, we discuss our results in the light of particle acceleration theory.


\section{The Model}

\label{sec:model}

To model the observational data we use the well-established
synchrotron self-compton (SSC) model \citep[e.g.][]{maraschi92}. We
assume a spherical, homogeneous emission region - coined blob -
containing isotropically distributed non-thermal electrons and a
randomly oriented magnetic field. \\Due to the presence of this
magnetic field the electrons emit synchrotron radiation. The photons
are then scattered off the same electron population via the inverse Compton
process. The resulting spectrum shows the typical two bump structure
commonly found in blazars. \\
In the following section the governing equations of the SSC-model are
explained.

\subsection{Photon distribution}

To determine the time-dependent spectral energy distribution of blazars we
solve the differential equation for the differential photon number density,
obtained from  the radiative transfer equation,
including the corresponding terms with respect to SSC model,

\begin{equation}
\frac{\partial n_{ph}(\nu)}{\partial t}	 =
R_S(\nu)-R_{SSA}(\nu)+R_{C}(\nu)-\frac{n_{ph}(\nu)}{T_{esc}}.
\label{eq:phot}
\end{equation}

\subsubsection{Synchrotron radiation}

In the following context the well known $\delta$-approximation \citep{fm66,
schlick02} is
applied to describe the synchrotron radiation in a convenient way. Thus
the synchrotron photon production rate $R_S$ is given by

\begin{equation}
R_S(\nu) = \frac{8 m c n_{e-}(\gamma_c) P_S(\gamma_c) }{3 e B h \nu \gamma_c},
\end{equation}
with the pitch angle averaged total power $P_S$ emitted by a single electron
having Lorentz factor $\gamma$ \citep{rybicki79, bg70, gs69}

\begin{equation}
P_S(\gamma)= \frac{4\; e^4 B^2 (\gamma^2-1)}{9\; m^2 c^3}.
\label{eq:PS}
\end{equation}
and   $\gamma_c$ being a function of $\nu$,

\begin{equation}
\gamma_c(\nu)= \sqrt{\frac{16 m c \nu }{3 e B}},
\end{equation}
obtained from the pitch angle averaged critical synchrotron frequency.


\subsubsection{Synchrotron-self-absorption}

In optically thick regimes the emitted synchrotron radiation is absorbed by the
emitting electrons itself. This is described by the synchrotron self absorption
coefficient,

\begin{equation}
\epsilon_{\nu}=-\frac{1}{12} \frac{c}{\nu^2 e B} \gamma_c P_S(\gamma_c)
\frac{\partial}{\partial \gamma}\left[
\frac{n_{e-}(\gamma)}{\gamma^2}\right]_{\gamma_c}  ,
\label{ssa}
\end{equation}
which leads to the absorption rate

\begin{equation}
R_{SSA}(\nu) =  c\;\epsilon_{\nu}\;n_{ph}(\nu).
\end{equation}

\subsubsection{Compton scattering}

The second main feature of the SSC model is Compton scattering of the
synchrotron photons by the emitting electrons themselves. Here the full
Klein-Nishina cross section is used to calculate the  photon production
rate,
\begin{align}
R_{C}(\nu) &= \int d\gamma\, n_{e}(\gamma) \; \times \nonumber
\\ &\times \int d\epsilon_1
\left[n_{ph}(\epsilon_1)\frac{dN(\gamma,\epsilon_1)}{dtd\epsilon}- n_{ph}(\epsilon)
\frac{dN(\gamma,\epsilon)}{dtd\epsilon_1}\right].
\end{align}
The formula was taken from \citet{wax05} with minor corrections according to
\citet{cb90}. The photon energies are rewritten in terms of the electrons rest
mass, so that $h \nu = \epsilon m c^2$ for the  scattered photons and  $h \nu =
\epsilon_1 m c^2$ for the target photons. To make use of the full Klein-Nishina
cross section we applied the approximate inverse Compton spectrum
\citep{jones68} of a single electron scattered off by a unit density photon
field,
\begin{align}
\frac{dN(\gamma,\epsilon_1)}{dtd\epsilon}& = \frac{2 \pi r_0^2 c}{\epsilon_1
\gamma^2} \left[ \frac{}{}2 q''\ln q''+(1+2q'')(1-q'')+\right.\nonumber
\\&+\left.\frac{1}{2} \frac{4 \epsilon_1 \gamma q'')^2}{(1+4\epsilon_1 \gamma
q'')}(1-q'')  \right],
\end{align}
where $ q''={\epsilon}/(4\epsilon_1 \gamma^2(1-\epsilon/\gamma))$ and
$1/(4\gamma^2)<q''\leq 1$. This equation is valid \\for \mbox{ $\epsilon_1<\epsilon\leq
4\epsilon_1\gamma^2/(1+4\epsilon_1\gamma)$}. The corresponding ordinary Compton
spectrum is approximately given by

\begin{equation}
\frac{dN(\gamma,\epsilon_1)}{dtd\epsilon}\approx \frac{\pi r_0^2 c}{2 \gamma^4
\epsilon_1}\left[(q'-1)(1+\frac{2}{q'})-2 \ln q'\right], \label{eq:345}
\end{equation}
with $q'= 4 \gamma^2 \epsilon/\epsilon_1$ and target photon energies in the range
$\epsilon_1/4\gamma^2\leq \epsilon < \epsilon_1$.

\subsubsection{Photon escape}

The last term describing the evolution of the photon number density represents
the photons escape rate. Here the photon escape time $T_{esc}$ is given by the
light crossing time
\begin{equation}
T_{esc}\approx\frac{R_b}{c},
\end{equation}
where $R_b$ is the radius of the emitting blob. The escape time is
chosen to be the light crossing time of the photons.

\subsection{Electron distribution}
The time evolution of the electron distribution is described by the kinetic equation
\begin{equation}
\frac{\partial n_{e^-}(\gamma)}{\partial t} = \frac{\partial}{\partial \gamma}\left( \frac{}{} n_{e^-}(\gamma)\; (\dot \gamma_S + \dot \gamma_{IC}) \right)-\frac{n_{e^-}(\gamma)}{T_{esc,e^-}}+Q_{inj}(\gamma).
\label{eq:elec}
\end{equation}
The synchrotron loss is given by $\dot \gamma_S=P_S(\gamma)/mc^2$ with the synchrotron power $P_S(\gamma)$ (cf. (\ref{eq:PS})). $T_{esc,e^-}= \eta R_b/c$ describes the electrons escaping from the emission region, where $\eta$ is an empirical factor. The inverse compton losses  $\dot \gamma_{IC}$ including the full Klein-Nishina cross-section are adopted following \citet{schlick02}. 
\begin{align}
\dot \gamma_{IC}&=\frac{3 \sigma_T c}{4}\int_0^\infty d \epsilon_1 \epsilon_1^{-1} n_{ph} (\epsilon_1)\\~
&\times \int_0^1 d q \frac{\Gamma_e^2}{(1+\Gamma_e q)^3} G(q,\Gamma_e)\nonumber
\end{align}
with
\begin{align}
G(q,\Gamma_e)= \left [ 2 q \ln q +(1+2q)(1-q)+\frac{(\Gamma_eq)^2(1-q)}{2(1+\Gamma_eq)}\right],\\
\Gamma_e=4\epsilon_1\gamma/(mc^2),\, q=\epsilon/[\Gamma_e(\gamma m c^2-\epsilon)]
\end{align}

%
%
%
%

As an injection function $Q_{inj}$ we use a power law combined with an exponential cut-off:
\begin{equation}
 Q_{inj}(\gamma)= K \gamma^{-s} \exp \left(\frac{-\gamma}{\gamma_{max}}\right),
\end{equation}
with Lorentz factor $\gamma$, the normalisation constant $K$,  cut-off energy
 $\gamma_{max}$ and the spectral index $s$. Using a injection function constant in time one yields a equilibrium solution
 for $n_{e^-}(\gamma)$ being constant in time.

\subsection{Numerics}
To obtain a model SED using the SSC formalism we solve the coupled equations \mbox{eq.
  (\ref{eq:phot})} and \mbox{eq. (\ref{eq:elec})}  numerically in our code framework. With respect to
stability issues we use the Crank-Nicholson scheme \citep{press02} to
compute the synchrotron part of the right hand side of both electrons equation and photons equation.
 The code was tested carefully and stands the
challenge of computing the equations in a range of 20 orders of
magnitude. All single effects (synchrotron radiation/losses, Compton
   scattering/losses) have been cross checked with analytical solution
   and approximation as well with numerical integration cross checks
   using Mathematica. Also comparison with existing codes were done
   successfully (as long as the models themselves were comparable).

\section{Results}
\label{sec:results}

\begin{figure*}
\centering
\includegraphics[width=\textwidth]{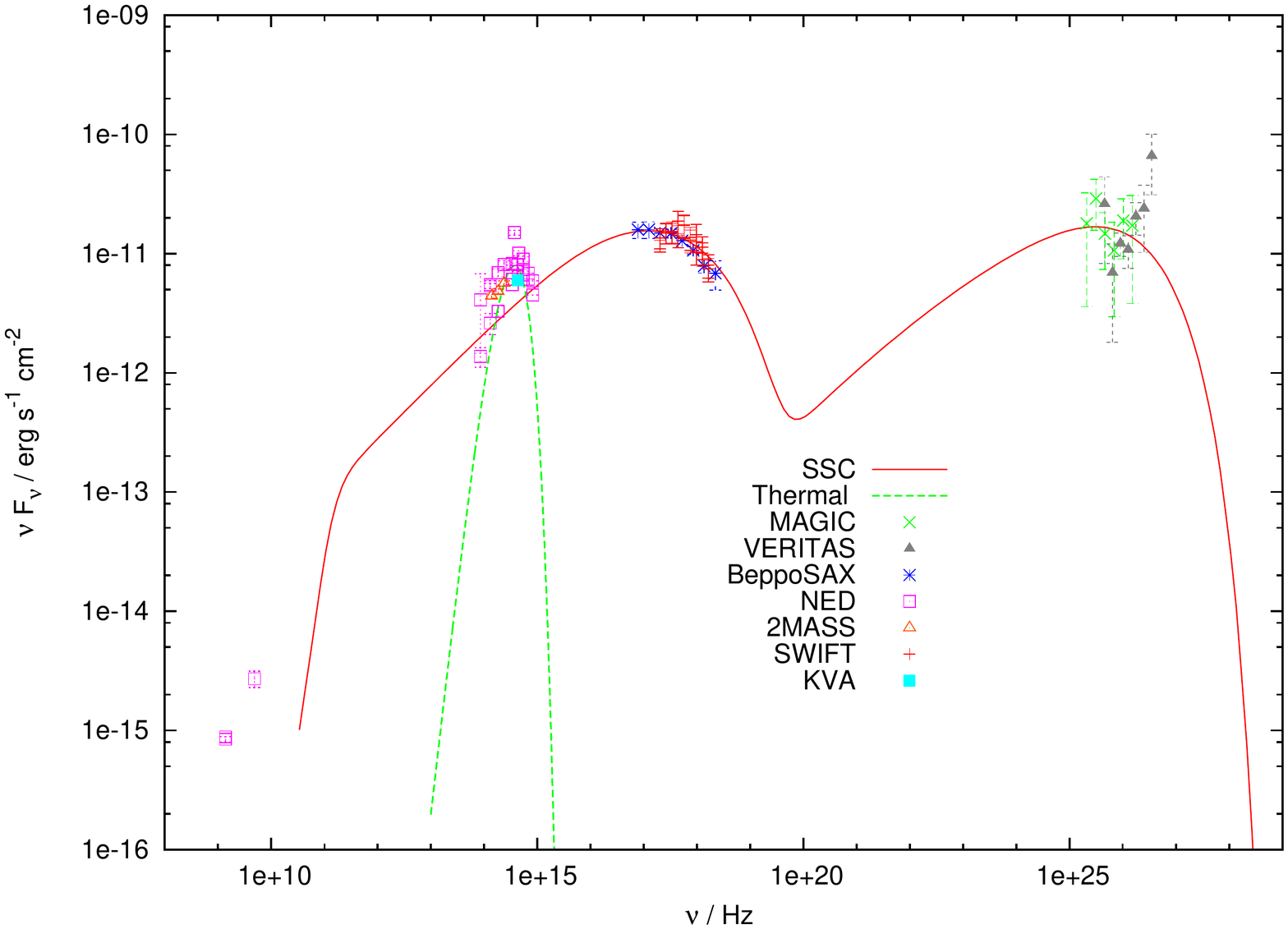}\caption{Overall SED of \source. Green crosses show the MAGIC data
\citep{albert06}, grey triangles the VERITAS data \citep{veritas09}. The TeV data have been de-absorbed by applying 
the EBL correction model of \citet{kneiske04}. The filled cyan box represents the simultaneously obtained KVA data
point. In the X-ray range SWIFT \citep{swift07} and BeppoSAX \citep{beppo05}
data are plotted as blue stars and red crosses. The 2MASS data (red open
triangles, \citep{twomass05}) together with the NED data represent the peak in
the optical range fitted by a thermal spectrum (dashed green line). The red
solid line depicts the SSC model curve.
 }
\label{fig:1}
\end{figure*}

In Fig.\ref{fig:1} we show the results of the application of the code to the
data of the HBL  \mbox{1 ES 1218+30.4}. Here we present a SSC model curve which
fits the data in the X-ray and VHE regimes. The  corresponding parameters are
listed in Tab. \ref{tab:1}. With this values we end up in an equilibrium state with a cooling break energy of the electrons distribution $\gamma_{break}= 6 \ttt 4$. 

The VHE data shown here have been discovered by the MAGIC telescope \citep{albert06}. Lately the VERITAS telescope could confirm this detection. As shown in \citet{veritas09}  the measured flux matches each other in the overlapping energy regime.

\setlength{\tabcolsep}{2pt}

\begin{center}
 \begin{table}
\begin{center}
 \begin{tabular}{ccccccc}
\hline\hline\\
$\gamma_{max}$      & $s$   & $B$/G   & $K/ \rm  cm^{-3}\;s^{-1}$ & $R$/cm       & $\delta$  & $\eta$ \\ \hline \\
$ 5.0 \ttt 5$       &$2.1$  & $0.04 $ & $0.4\ttt {-1} $     &$3 \ttt {15} $& $80 $     & $10$ \\
\hline\hline
\end{tabular}
\caption{Best fit parameters for the SSC modelling the SED of 1 ES 1218+30.4}
 \label{tab:1}
\end{center}
\end{table}
\end{center}
The data in the X-ray regime have been taken by SWIFT between March
and December 2005 \citep{swift07}. Another set of X-ray data were
obtained by the BeppoSAX experiment in 1999 \citep{beppo05}. It is
remarkable, that in spite of a lag of 6 years between these
observations there is no difference in the X-ray flux. Together with 
the constant TeV flux level this is a strong argument for an almost constant 
background of non-thermal electrons in a constant magnetic field.

The KVA data point shown in Fig. \ref{fig:1} was obtained simultaneously to the
MAGIC observation of \mbox{1 ES 1218+30.4}. One can also see that the SSC model is
not able to fit this data point. This discrepancy can be resolved by taking the
NED and 2MASS data \citep{twomass05} surrounding the optical KVA point into
account.
This set of data points has been modeled by a simple blackbody spectral
distribution with a temperature $T=  4500$K and a radius \mbox{$R_{bb} = 2.85
\ttt {16} $cm}.  Considering average sun-like stars we yield an estimate
of the number of stars responsible for this blackbody radiation given by
$R_{bb}^2/R_{\sun}^2=1.7 \ttt {11}$. This number together with the temperature
$T$ gives a hint for the host galaxy of \mbox{1 ES 1218+30.4} being the origin
of this feature in the SED not being modelled by the SSC approach. With this blackbody approximation 
we get an estimate for the central black hole mass of the AGN using the correlation given by \citet{kormendy01}
\begin{equation}
M_{BH}=0.78\ttt8 M_{\sun} \left(\frac{L_{bulge}}{10^{10} \, L_{\sun}} \right).
\end{equation}
Here we get $M_{BH}=5.6\ttt8$ and a corresponding Eddington luminosity $L_{Edd}=7.3\ttt{46}$. 
Lacking observations of the radial dependence
of the surface brightness of the host galaxy, the accuracy
of these estimates is not better than a factor of a few.
Adopting a bulk Lorentz factor $\Gamma=57$ ,
which is consistent with  $\delta=80$ for reasonable angles of the jet axis to the
line-of-sight, the injected luminosity 
in the AGN frame is  $L_{inj}= 4/3 \pi R_b^3 \Gamma^2 \int d\gamma \,\gamma m\,c^2\,Q(\gamma) =  7.3\ttt{43}$ erg/s.
 The resulting very low Eddington
ratio of the order of 0.001 is in line with the results of population studies
of BL Lac objects, for which \citet{treves02} find 0.01.

%


\section{Discussion}

In this paper, we presented SSC model fits to the contemporaneous data of
\source in the X-ray and VHE range. In the optical regime
a simple blackbody spectrum has been applied to complete the
model SED. In a similar way, \citet{kat01} modeled
the optical data of Markarian 501 by applying a standard model for
the elliptical host galaxy  \citep{nil99}. Our simple approach
to the background radiation of the host galaxy actually suffices, the
more sophisticated approach of \citet{kat01} might be
of use if more data for the host galaxy would be available, but at
this high redshift details are not accessible. \citet{celotti08} 
have recently studied \source, assuming it is in a flaring state.

Observational results for 1 ES 1218+30.4 have also been
discussed in \citet{sato08} with special regard to the variability
of the source. 
The authors concluded that the source must
have a very hard electron distribution with power law slope s = 1.7. We disagree
with this result, as we were able to show that there exist model parameters
which are well in line with relativistic shock acceleration
theory \citep{ellison90}, although harder spectra can be imagined
for more extreme sources \citep{vs98}.

The major difference of our spectrum compared to \citet{sato08} is that
we assumed the optical regime to be dominated by the
host galaxy, approximately described by a blackbody spectrum.
Therefore, the need for extreme electron spectra could be
relaxed.  Considering that the SWIFT, MAGIC and
VERITAS data used here do not show strong flaring features, we have modeled
the SED as steady-state SSC emission with our time-dependent code,
obtaining physical parameters of the emission region.  These parameters
lie well in the range found with SSC models for other HBLs. The small magnetic field value differs slightly from the normally used values of about $0.1 - 0.2$ G in SSC models, but in contrast to competing hadronic models this value is still reasonable.

Additionally, the central black hole mass could be estimated from the host
galaxy properties,
demonstrating that the emission region is consistent with a
sub-Eddington jet as generally expected for BL Lac-type sources.The growing, but
still marginal, 
discrepancy of our model SED and the VHE spectra at highest energies,
if taking the expected gamma ray attenuation due to pair production into
account, is a concern. 
It is also found in other studies of TeV blazars.  This trend could indicate an
insufficiency of the SSC model approach, a weaker than expected
gamma ray attenuation, or an incomplete understanding of the energy
determination of air showers from their Cherenkov emission.

\section*{Acknowledgements}

 MR acknowledges support from the Deutsche Forschungsgemeinschaft by
  Graduiertenkolleg 1147, FS acknowledges support from the Deutsche
  Forschungsgemeinschaft through grant SP 1124/1

\bibliographystyle{mn2e} 
\bibliography{apj-jour,1218}

\label{lastpage}

\end{document}